\documentclass[aps,prl,twocolumn,showpacs,letter,superscriptaddress,amsmath,amssymb]{revtex4}
\usepackage[dvips]{graphicx}
\DeclareGraphicsExtensions{.eps,.pdf}
\usepackage{setspace} 
\pdfoutput=1

\usepackage{dcolumn}
\usepackage{bm}
\usepackage{soul}
\usepackage[latin1]{inputenc}
\usepackage{multirow}
\usepackage{color} 
\usepackage{bm}	
\newcommand{\expo}[1]{\text{e}^{#1}}		

\begin{document}

\title{Macroscopic effects in attosecond pulse generation}

\author{T. Ruchon}
\affiliation{Department of Physics, Lund University, P.
O. Box 118, SE-221 00 Lund, Sweden}

\author{C. P. Hauri}
\affiliation{Laboratoire d'Optique Appliqu\'ee, \'Ecole Nationale Sup\'erieure des Techniques Avanc\'ees (ENSTA) - \'Ecole Polytechnique CNRS UMR 7639, 91761 Palaiseau, France}

\author{K. Varj\'u}
\affiliation{Department of Physics, Lund University,
P. O. Box 118, SE-221 00 Lund, Sweden}

\author{E. Gustafsson}
\affiliation{Department of Physics, Lund University, P.
O. Box 118, SE-221 00 Lund, Sweden}

\author{R. L\'opez-Martens}
\affiliation{Laboratoire d'Optique Appliqu\'ee, \'Ecole Nationale Sup\'erieure des Techniques Avanc\'ees (ENSTA) - \'Ecole Polytechnique CNRS UMR 7639, 91761 Palaiseau, France}

\author{A. L'Huillier}
\affiliation{Department of Physics, Lund University,
P. O. Box 118, SE-221 00 Lund, Sweden}

\begin{abstract}
We examine how the generation and propagation of high-order harmonics in a partly ionized gas medium affect their strength and  synchronization. The temporal properties of the resulting attosecond pulses generated in long gas targets can be significantly influenced by macroscopic effects, in particular by the intensity in the medium and the degree of ionization. Under some conditions, the use of gas targets longer than the absorption length can lead to the generation of self-compressed attosecond pulses. We show this effect experimentally, using long argon-filled gas cells as generating medium.\end{abstract}
\pacs{42.65Ky, 42.50Hz}

\maketitle 
The generation of attosecond pulses from high-order harmonic conversion in gases requires that harmonics selected in a given bandwidth be synchronized\,\cite{PaulScience2001}. In general, this condition is only realized naturally in the cut-off region\,\cite{KienbergerNature2004,MairessePRL2004}. In the plateau region the variation of the harmonic spectral phase tends to ruin the attosecond structure. Indeed, interferences between contributions from different quantum paths responsible for the harmonic emission\,\cite{AntoinePRL1996,GaardePRL2002} lead to multiple pulses every half laser cycle while the variation of the accumulated phase along a given quantum path from one harmonic to the next results in a significant temporal chirp\,\cite{MairesseScience2003}. To compensate for the multiple pulse structure, the short quantum path is usually selected by phase-matching or spatial filtering\,\cite{SalieresScience2001,LopezMartensPRL2005}. To eliminate the phase variation due to the different excursion times for the short quantum path, post-compression techniques have been proposed, based on broadband multilayer mirrors\,\cite{MorlensOL2006}, metallic filters\,\cite{LopezMartensPRL2005,KimPRA2004,SansoneScience2006} or plasma media\,\cite{MairesseScience2003}. These techniques however reduce the number of photons available, and must be designed specifically for each spectral region.

For the generation of attosecond pulses in non-linear media, both the individual single-atom response and phase matching effects must be taken into account.  These effects are naturally included in complete calculations solving the Schrödinger and Maxwell equations\,\cite{GaardePRL2002,GaardeOL2006,ChristovPRA1998}. Besides, the optimization of individual harmonics has required careful study of phase matching conditions \,\cite{LHuillierJPB1991,ConstantPRL1999,KazamiasPRL2003,zhangnatphysics2007}. However, the influence of these macroscopic effects on the relative spectral phase of several consecutive harmonics, and consequently on the temporal shape of attosecond pulses has not so far been scrutinized. In fact, the remarkable agreement between measurements of spectral phases and the predictions coming from the single-atom dipole phase\,\cite{MairesseScience2003} seems to indicate that the macroscopic response should not influence the temporal properties of attosecond pulses, at least under the experimental conditions hitherto investigated. 

In this Letter, we theoretically examine the influence of the macroscopic response on the temporal properties of the attosecond pulses. For targets with lengths longer than the coherence and absorption lengths of the generation process, we show how dispersion in neutral and ionized media indeed influences the spectral phase of attosecond pulses through phase-matching effects. We demonstrate this effect experimentally, by measuring the harmonic group delay for different laser focus positions in a 10 cm-long argon-filled gas cell. The shortest and cleanest pulses are obtained when the intensity in the region where harmonics are generated is the lowest, in contradiction with single atom predictions.

We consider a coherent sum of consecutive odd harmonics (from $q_i$ to $q_f$) generated when a laser field interacts with a gas. The electric field resulting from the superposition of these harmonics can be written as 
${\cal E}\left( t,z'\right)=\sum_{q=q_i}^{q_f} {\cal E}_q(t,z')$, 
where ${\cal E} _q$ is the electric field of the $q$-th harmonic, $t$ is time and $z'$ is the abscissa of the observation point, supposed to be far away from the generating medium (see Fig.~1(a)). Using an integral formalism, ${\cal E}_q(t,z^\prime)$ can be simply calculated by integrating the nonlinear polarization, ${\cal P} _q$, induced in the medium at position $z$ and at time $t-\ell_q(z,z^\prime)/c$, where $\ell_q(z,z^\prime)$ is the optical path of harmonic $q$ from $z$ to $z^\prime$. In a one-dimensional approximation, valid for loose focusing geometries, 
\begin{equation}
{\cal E}_q\left(t,z^\prime\right)\propto \int {\cal P}_q\left(t-\frac{\ell_q\left(z,z^\prime\right)}{c},z\right) dz.
\label{phasemat}
\end{equation}
${\cal P}_q(t,z)\propto {\cal N} d_q \exp{iq\omega(t-\ell_1(-\infty,z)/c)}$  
where $d_q$ is the single atom dipole moment, $\cal{N}$ the atomic density, $\omega$ the laser frequency and $\ell_1\left(-\infty,z\right)$ is the optical path of the fundamental from $-\infty$ till $z$. Introducing the complex-valued wave vectors $k_1$ and $k_q$,    
\begin{equation}
{\cal E}_q\left(t,z^\prime\right)\propto\expo{i q\omega t}\int {\cal N} d_q \expo{-iq\int_{-\infty}^z k_1dz^{\prime\prime}-i\int_z^{z^{\prime}} k_q dz^{\prime\prime} } dz.
\label{Efield}
\end{equation}
To gain some physical insight into this equation, we consider for simplicity a homogeneous medium of length $L$ and a collimated geometry.  Absorption of the fundamental is considered to be extremely small and not taken into account, while absorption at the frequency of the $q$th harmonic, denoted $\kappa_q$, is described by the imaginary part of $k_q$. In the following, we introduce the phase mismatch $\Delta k_q$ equal to the real part of $k_q-qk_1$. Eq.\,\eqref{Efield} becomes
\begin{equation}
{\cal E}_q (t, z^\prime)\propto \expo{i q\omega (t-{\ell_1\left(-\infty,z'\right)}/{c})}\; {\cal N}d_q \frac{1-\expo{(-i\Delta k_q -\kappa_q) L}}{i\Delta k_q+\kappa_q }
\label{phasemat2}
\end{equation}
The amplitude of the electric field at frequency $q\omega$ depends on two complex quantities: $d_q= |d_q| \exp{i\phi^{mic}_q}$ describes the single atom response, while 
\begin{equation}
F_q= \frac{1-\expo{(-i\Delta k_q -\kappa_q) L}}{i\Delta k_q+\kappa_q }=|F_q|\expo{i\phi^{mac}_q}
\end{equation}
includes macroscopic effects such as phase matching, absorption of the neutral medium and effects due to ionization. $\phi^{mac}_q$ simplifies in two limiting cases: when $L$ is much larger than the absorption lengths ($L_q^{abs}=\kappa_q^{-1}$), $\phi^{mac}_q \approx -\arctan(\Delta k_q L_q^{abs})$. The influence of dispersion over the spectral phase is here limited to a length equal to the absorption length, which is physically quite intuitive. When $L$ is much smaller than both $L_q^{abs}$ and $L_q^{coh}=\Delta k_q^{-1}$, $\phi^{mac}_q \approx -\Delta k_q L/2$. Again, this makes sense since in absence of absorption the generated harmonic will propagate on average half the medium length.

 In the left panel of Fig.~1(b), we show the variation of $\Delta k_q$ due to neutral atom dispersion \cite{LhuillierJOSAB1990} in a 20 mbar Ar target (blue circles) and due to free electron dispersion assuming 7\% ionization of the medium (green squares), over a (continuous) frequency range spanning from the 11th to the 30th harmonic of 800 nm radiation. This degree of ionization in argon corresponds to optimized phase matching conditions where the dispersion of neutral atoms cancels that of free electrons for most of the bandwidth considered (see below). The total phase mismatch is indicated by the red line and red symbols. The right panel of Fig.~1(b) shows $\phi^{mic}_q$ \cite{varjuJMO2005} (blue circles) and $\phi^{mac}_q$ (green squares) for a $L=3$ mm medium. We also indicate $\phi^{mac}_q$ for $L\gg L_q^{abs}$ (red diamonds). The nonlinear variation of $\phi_q^{mac}$ with respect to frequency will in general influence the temporal properties of the attosecond pulses. In the case discussed in Fig.~1, its curvature is opposite to the spectral phase variation due to the single atom response. It will therefore compensate it, leading to shorter attosecond pulses. 

In Fig.\,\ref{figtheo2} (a-b), we plot the phase matching factor $|F_q|$ (a) and the macroscopic group delay (GD) $GD_q^{mac}=\partial \phi_q^{mac}/ \partial q\omega$ (b) versus the length of the cell and the ``harmonic order", $q$, for the same conditions as in Fig.\,\ref{figtheo1}\,(b, left), i.e. 20 mbar Ar, with 7\% ionization. The GD decreases with $q$ especially for long medium lengths. Since the single atom GD increases with $q$ for the short quantum path \cite{MairesseScience2003}, both effects will compensate each other leading to better harmonic synchronization. In the case where phase matching is not optimized, for example, for higher ionization rates (14 \%), both $|F_q|$ (c) and $GD_q^{mac}$ (d) exhibit pronounced oscillations as a function of medium length. The amplitude oscillations, well known in nonlinear optics under the name of Maker fringes, result from the fact that interferences between harmonic fields generated in different parts of the medium strongly depend on the length of the medium. As expected, the GD exhibits similar oscillations. We here envision possibilities to control the attosecond pulse duration in conditions where individual harmonics are not perfectly phase matched.

Figure\,\ref{figtheo3} presents the temporal properties (pulse duration and contrast) of attosecond pulses obtained by superimposing harmonics 11 to 31 for cell lengths varying from 0 to 20\,mm and ionization rates from 0 to 20\,\%. We here included the microscopic response by using the dipole moment intensities calculated by integrating the time-dependent Schr\"odinger equation in argon~\cite{GaardePRA2000} and dipole phases $\phi^{mic}_q$ characteristic of the shortest quantum path~\cite{LewensteinPRA1995b}. To give a better description of the temporal profile than the full width at half maximum, we use the second order moment defined as 
\begin{equation}
\bar{\tau}= \left[\frac{1}{W} \int_{-\infty}^{\infty} t^2I(t)dt -\frac{1}{W^2}\left( \int_{-\infty}^{\infty} tI(t)dt\right)^2\right]^{1/2}
\end{equation}
where $W$ is the integral of the intensity profile $I(t)$, i.e. the energy fluence. Second order moments are equal to 42\,\% of the full width at half maximum for Gaussian pulses and increase rapidly as the contrast of the pulses decreases.  The main effect shown in Fig.~\ref{figtheo3} is that there is a region of medium lengths and ionization rates where the temporal profiles of the generated attosecond pulses are optimized. This happens for lengths larger than the average absorption length ($L_{23}^{abs}$ is indicated by the red dashed line) and for ionization rates in the range 5 to 7\%. The white dashed lines show the coherence length (abscissa) for the 23rd harmonic as a function of ionization rate (ordinate). The shortest pulses are here obtained for optimized phase matching conditions where the coherence length is going to infinity and for medium lengths typically longer than twice the absorption length. We also verified that more refined calculations taking into account the (weak) $z$-dependence of the dipole moment and phase as well as the Gouy phase did not significantly change our results.  

To show experimentally the influence of macroscopic effects on the spectral phase of attosecond pulses, we generated harmonics in a long cell, using a loose focusing geometry and varied the position of the focus in the generating cell, as illustrated in Fig.~\ref{figexp}(a). The 1\,kHz titanium sapphire laser of the Lund High-Power Laser Facility, with 1.4\,mJ pulse energy, 6\,mm beam diameter and 35\,fs pulse duration was focused by a 1 m lens to an intensity of $1.6 \times 10^{14}\text{W}/\text{cm}^{2}$. The gas cell was a 10\,cm long glass tube, with an inner diameter of 800\,$\mu$m. Ar gas was pulsed at 500\,Hz through a hole drilled in the middle of the cell with a piezo-electric valve. The beam waist was estimated to $\approx 100 \mu$m, i.e. roughly eight times smaller than the tube diameter, thus avoiding any guiding by the tube. We could observe the blue plasma light emission characteristic of argon in almost the whole target when focusing at its center, which substantiated the fact that we had both the intensity and the matching pressure in the whole target to partially ionize the medium. Finally, a hard aperture allowed us to filter the short quantum path contribution to the high harmonic emission~\cite{LopezMartensPRL2005} and a 200 nm-thick Al filter was used to eliminate the fundamental laser field.

Our attosecond pulses were characterized using the RABITT method (Reconstruction of Attosecond Beating by Interferences of Two-photon Transitions)~\cite{PaulScience2001}. The laser beam was split in two arms to perform an XUV+IR pump-probe cross correlation. The main fraction of the pulse energy went into the pump arm where attosecond pulses were generated, as described above. The small part of the laser beam, sent into the delayed probe
arm, was recombined with the XUV pump immediately after the spatial filter, using the coated surface of the aperture to
reflect the probe beam. The recombined beams were then focused into a magnetic bottle electron spectrometer, filled with argon at a detection pressure of $1\; \text{to}\;10\times 10^{-4}$~mbar. By taking into account the phase and amplitude effect
of ionization of the detection gas and the Al filter, we could get the spectrum and the
GD of the generated attosecond pulses (Fig.\,\ref{figexp}(b)) and reconstruct their average temporal profile (see right panels in Fig.\,\ref{figexp}(b))\,\cite{VarjuLP2005}.

The overall shape of the harmonic spectra does not vary significantly as a function of the focus position, while the energy output is approximately four times smaller for $z_0=45$\,mm than for $z_0=10$\,mm. The GD on the other hand becomes flatter and flatter as $z_0$ moves towards the end of the medium. Accordingly the pulse contrast improves. Fig.\,\ref{figexp}(c) shows the second order moments as a function of focus position. As the focus moves toward the end of the cell, the second order moment, indicated by the green squares, decreases from 260 to 120 as. The intensity at the end of the gas cell where the harmonics are effectively generated (in the dashed region) decreases from 1.6$\times 10^{14}\,\text{W/cm}^2$ to 1.2$\times 10^{14}\,\text{W/cm}^2$ while the ionization rate is estimated to decrease from 23\,\% to 6\,\%. In these conditions, in contrast, the single atom response predicts a flatter GD and thus shorter pulses as the intensity increases (red line)\,\cite{MairessePRL2004}. The results of our simple model are indicated by the blue squares. We get a better agreement with the experimental results than if we only consider the single atom response. The remaining discrepancy might be due to the approximations inherent to our simple model which includes neither three dimensional effects nor time integration. Another possible effect is the reshaping of the fundamental while propagating in the long cell\,\cite{GaardeOL2006}. In spite of these approximations, our model predicts a deterioration of the temporal contrast as the intensity increases, in agreement with the experimental observations. It also gives some insight into the contribution of phase matching effects to attosecond temporal properties.


In conclusion, we have shown that phase matching plays a significant role in the temporal shaping of attosecond pulses. A theoretical description has been provided showing that getting intense and short attosecond pulses is a trade off between the length of the cell and the laser intensity which dictates the ionization rate. This work opens the way to the design of temporal quasi phase matched targets that could provide self-compressed attosecond pulses by modulating the ionization rate along the propagation axis~\cite{zhangnatphysics2007,KosugePRL2006}. 

\begin{acknowledgments}
This research was supported by 
the Integrated Initiative of Infrastructure
LASERLAB-EUROPE (RII3-CT-2003-506350) within the 6th European
Community Framework Programme, the Marie Curie Research
Training Network XTRA (MRTN-CT-2003-505138), 
the Swedish Research Council and the Knut and Alice Wallenberg
Foundation. T. R. is supported by a Marie Curie Intra-European fellowship (MEIF-CT-2006-040577).
The authors thank Bertrand Carr\'e and Claes-G\"oran Wahlstr\"om for fruitful discussions. 
\end{acknowledgments}
\newpage
\hyphenation{Post-Script Sprin-ger}

\newpage
\newpage
\section{Captions}
\subsection{figure 1}
\label{sec:figure1}
(a) Notations used for the theory. (b) (Left panel) Phase mismatch vs. harmonic order due to the electrons (green squares), the neutral atoms (blue circles) and their sum (red diamonds). (Right panel) Microscopic ($\phi^{mic}$, blue circles) and macroscopic phase ($\phi^{mac}$, green squares) versus harmonic order. The red diamonds indicate the macroscopic phase in the limit $L\gg L^{abs}$.
\subsection{figure 2}
\label{sec:figure 2}
Amplitude of the phase matching factor $|F_q|$ vs the length of the cell and the harmonic order for  (a) 7\% and (c) 14\% ionization.  (b) Macroscopic group delay, i.e. ${\partial \phi_q^{mac} }/{\partial q \omega}$ vs $L$ and $q$ for (b) 7\% and (d) 14\% ionization. The pressure was 20 mbar.
\subsection{figure 3}
\label{sec:figure 3}

Second order moments of the attosecond pulses reconstructed from the superposition of harmonics 11 to 31 for ionization ratio from 0 to 20\% and cell length from 0 to 20\,mm. The white (resp. red) dash line represents the coherence (resp. absorption) length of harmonic 23rd.
\subsection{figure 4}\label{sec:figure 4}
 (a) Experimental arrangement to vary the intensity of the generating laser field. The photons that exit the medium with no reabsorption are emitted in the last few millimeters of the target (dashed area). (b) (Left) Spectra and group delay measured for (top) $z_0 = 5$\,mm (higher intensity) and (bottom) $z_0 = 45$\,mm (lower intensity). (Right) Corresponding temporal profiles.  (c)  Experimental (green squares) second order moments for different focus positions. Theoretical second order moments taking into account the microscopic phase only (red line) and the total phase (blue squares). The theoretical curves are obtained assuming an intensity going from 1.6$\times 10^{14}$\,W/cm$^2$ ($z_0 = 5$\,mm) to 1.2$\times 10^{14}$\,W/cm$^2$ ($z_0 = 45$\,mm).

\begin{figure}[htbp]
		\begin{center}
\includegraphics[width=1.00\linewidth]{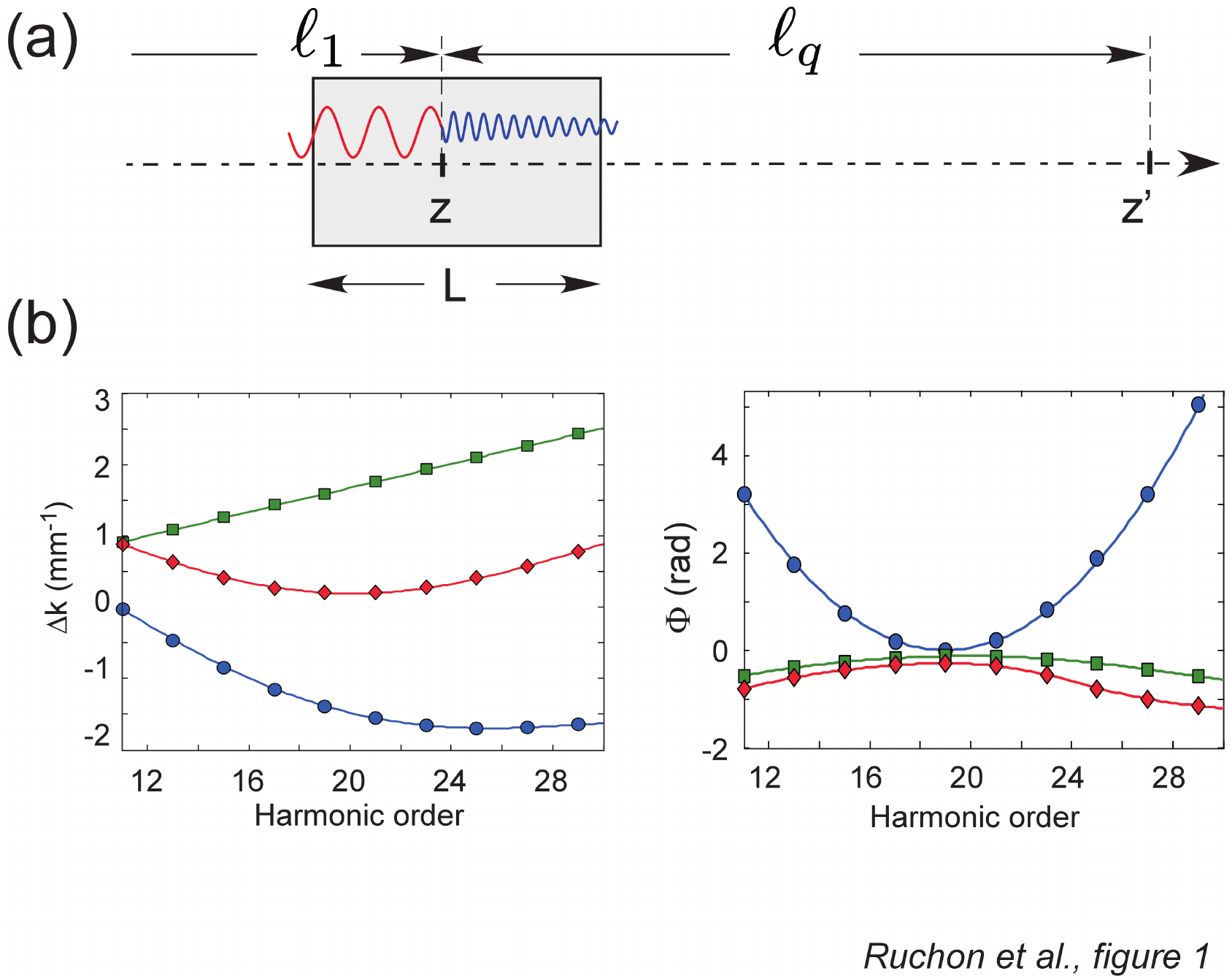}
	\end{center}
	\caption{	\label{figtheo1} }
\end{figure} 

\begin{figure}[htbp]
		\begin{center}
\includegraphics[width=1.00\linewidth]{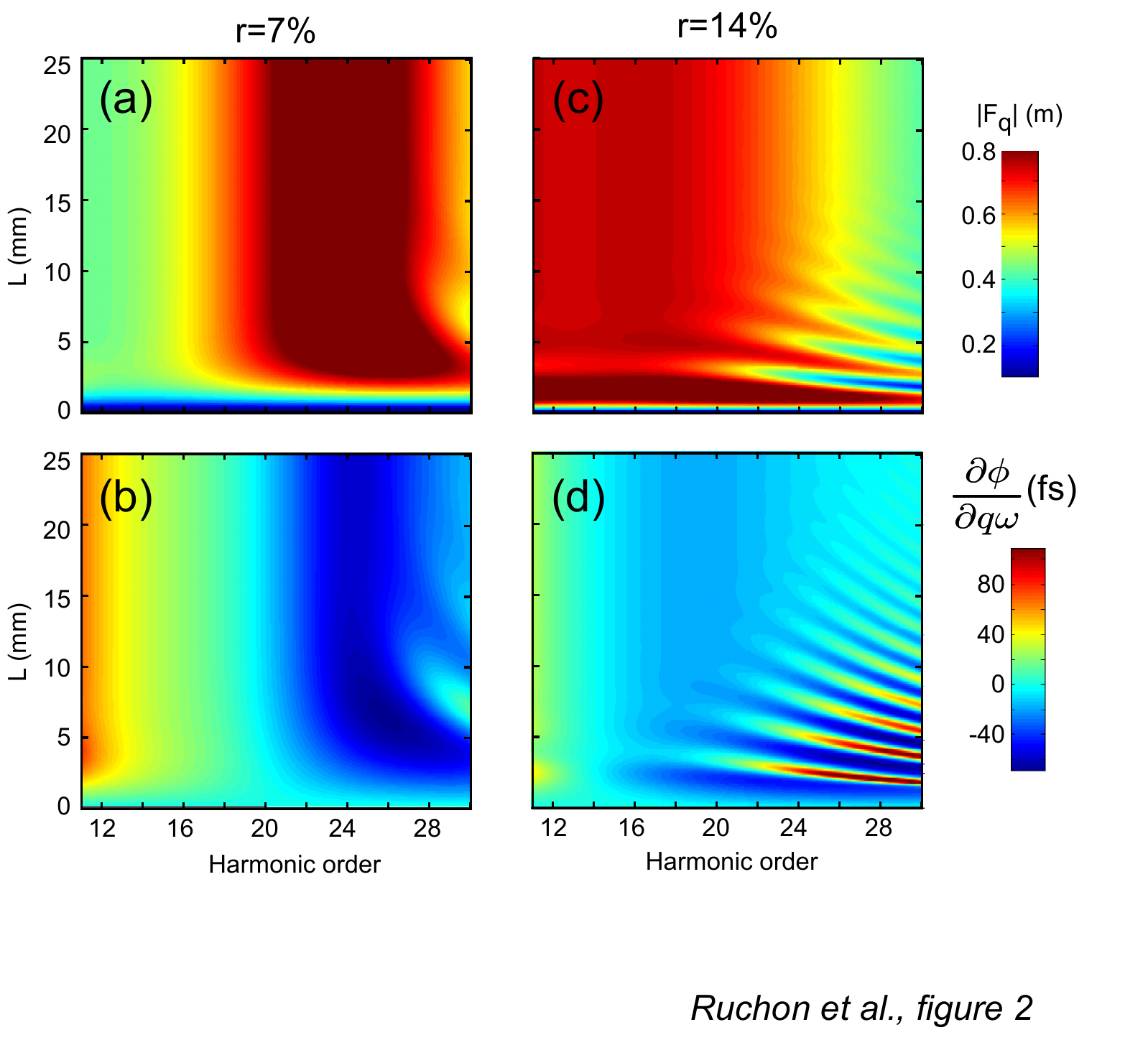}
	\caption{	\label{figtheo1} }
	\end{center}
\end{figure} 

	\begin{figure}[htbp]
		\begin{center}
\includegraphics[width=1.00\linewidth]{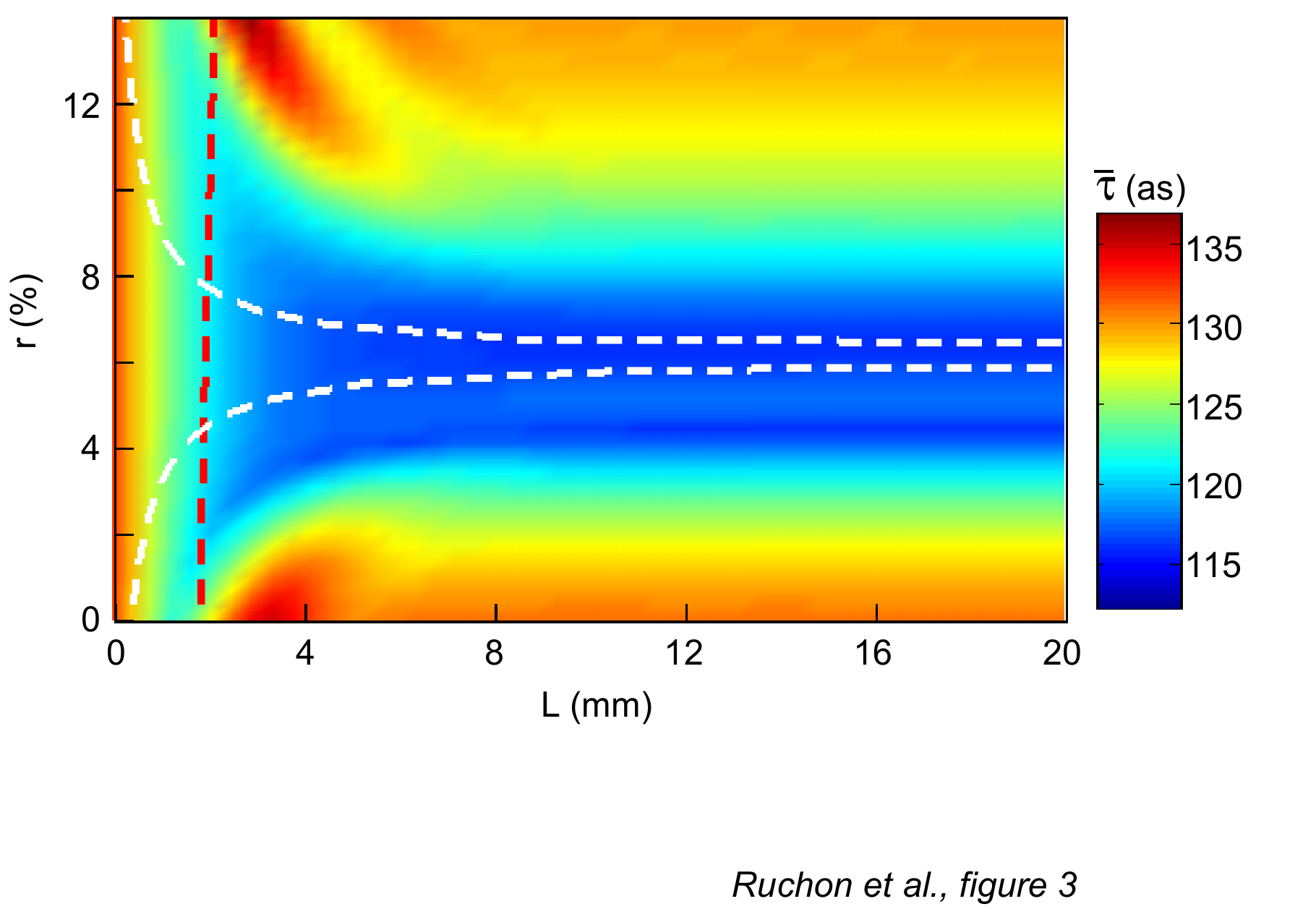}
	\end{center}
	\caption{	\label{figtheo3} }
\end{figure}
	\begin{figure}[htbp]
		\begin{center}
\includegraphics[width=0.95\linewidth]{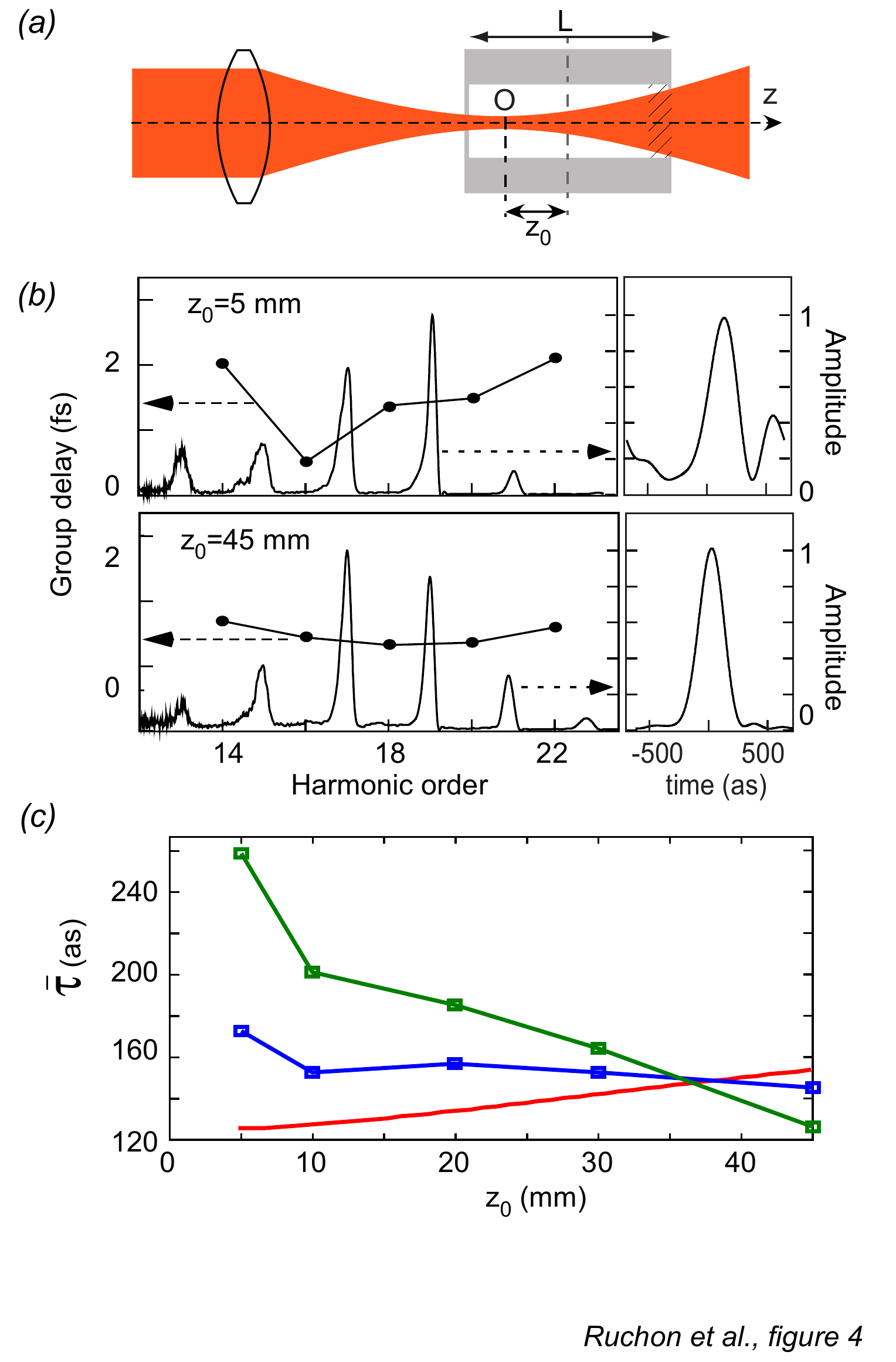}
	\end{center}
	\caption{	\label{figexp} (Color online)}
\end{figure}

\end{document}